\documentclass[twocolumn,superscriptaddress,aps,pra]{revtex4-2}

\makeatletter
\newcommand*{\rom}[1]{\expandafter\@slowromancap\romannumeral #1@}
\makeatother

\usepackage{color}
\usepackage{graphicx}
\usepackage{hyperref}
\usepackage{amsmath,amssymb}
\usepackage{epstopdf}
\usepackage{subcaption}
\usepackage{float}

\graphicspath{{figs/}}

\def\beq{\begin{equation}}
\def\eeq{\end{equation}}
\def\bea{\begin{eqnarray}}
\def\eea{\end{eqnarray}}

\usepackage{xcolor}

\begin{document}
\pagestyle{plain}

\title{Thermodynamic Geometry of Classical and Quantum Statistics in the Relativistic Regime}
\author{Hosein Mohammadzadeh}
\email{mohammadzadeh@uma.ac.ir}
\affiliation{Department of Physics, University of Mohaghegh Ardabili, P.O. Box 179, Ardabil, Iran}
\author{Zahra Ebadi}
\affiliation{Department of Physics, University of Mohaghegh Ardabili, P.O. Box 179, Ardabil, Iran}
\author{Omid Yahyayi}
\affiliation{Department of Physics, University of Mohaghegh Ardabili, P.O. Box 179, Ardabil, Iran}
\author{M. H. Naghizadeh Ardabili}
\affiliation{Department of Physics, University of Mohaghegh Ardabili, P.O. Box 179, Ardabil, Iran}

\begin{abstract}
We investigate the thermodynamic geometry of classical and quantum ideal gases in the relativistic regime, with particular emphasis on the effects of particle mass and spatial dimensionality. Relativistic kinematics is incorporated through the full energy-momentum dispersion relation and the corresponding relativistic density of states. Using the Fisher-Rao information metric derived from the partition function, we analyze the thermodynamic curvature for Maxwell-Boltzmann, Bose-Einstein, and Fermi-Dirac statistics. Exact analytical expressions are obtained in two spatial dimensions, while the three-dimensional case is studied numerically. We show that the thermodynamic curvature preserves its characteristic sign—positive for bosons and negative for fermions; even in the relativistic regime, reflecting effective attractive and repulsive statistical interactions, respectively. A distinctive relativistic effect is the shift of curvature singularities from the non-relativistic critical point to a mass-dependent threshold at $\mu = mc^{2}$. In addition, the relativistic Bose-Einstein condensation temperature is evaluated, revealing explicit mass-dependent corrections to the non-relativistic result. These findings provide a unified geometric perspective on relativistic statistical systems and clarify the interplay between quantum statistics, relativistic kinematics, and critical behavior.
\end{abstract}

\maketitle
\section{Introduction}\label{1}

The foundational distributions of statistical mechanics—namely the classical Maxwell–Boltzmann (MB) and the quantum Bose–Einstein (BE) and Fermi–Dirac (FD) statistics—play a central role in describing the equilibrium properties of many–body systems ranging from condensed–matter physics to cosmology. These distributions arise from the intrinsic quantum nature of particles: bosons, obeying symmetric wavefunctions and unrestricted occupancy, and fermions, characterized by antisymmetric states and constrained by the Pauli exclusion principle. This difference manifests through the commutation or anticommutation relations between creation and annihilation operators and ultimately determines the macroscopic behavior of the corresponding systems \cite{pathria2011statistical,huang1991statistical,niven2004exact}.

Although the MB, BE, and FD distributions in the non‑relativistic regime have been extensively studied and generalized, comparatively little attention has been devoted to their counterparts in the relativistic regime. In this work, we investigate these statistics from a geometric perspective by incorporating relativistic effects through both the density of states and the complete energy–momentum dispersion relation, with special emphasis on the influence of particle mass. The functional form of these distributions remains unchanged as long as the basic principles of particle counting and exchange symmetry are preserved; only the density of states is modified due to the relativistic nature of the energy spectrum. Nevertheless, in curved space‑times, where the local definition of temperature becomes relevant, the form of the distribution function itself may vary—an aspect that lies beyond the scope of the present study \cite{rovelli2013general,KowalskiGlikman2021,Tolman1934,Tercas2013}.

Thermodynamic geometry has proven to be a powerful and insightful framework for analyzing the microscopic structure of thermodynamic systems, especially those involving complex interactions or correlated ensembles. Pioneered by Weinhold and Ruppeiner, this formalism introduces a Riemannian geometric structure on the thermodynamic parameters space, wherein the curvature of the associated metric encodes fundamental information regarding interactions and stability \cite{ruppeiner1979thermodynamics,weinhold1975metric,ruppeiner1995riemannian}.

Within this approach, the thermodynamic curvature—a scalar obtained from the Riemann curvature tensor—plays a key interpretative role. A positive curvature is commonly associated with attractive interactions, characteristic of bosonic systems, while a negative curvature indicates repulsive interactions, typical for fermionic systems. A vanishing curvature corresponds to an ideal, non–interacting gas, whereas curvature singularities often mark phase transitions and critical points where thermodynamic quantities exhibit non–analytic behavior \cite{ruppeiner1995riemannian,janyszek1990riemannian}.

Over recent decades, thermodynamic geometry has found broad application across diverse areas of physics, including black–hole thermodynamics, strongly correlated fermionic systems, and relativistic or cosmological models. Its geometric language provides an elegant bridge between microscopic statistical behavior and macroscopic thermodynamic observables—particularly useful for systems where standard perturbative methods become intractable. In relativistic gases, the intricate coupling between mass, temperature, and dimensionality gives rise to rich thermodynamic phenomenology, rendering a geometric analysis both natural and insightful for identifying stability and critical phenomena.

The relativistic regime introduces additional subtleties: the density of states acquires explicit dependence on the relativistic dispersion relation, and the scaling behavior of thermodynamic quantities changes significantly. The rest mass becomes a crucial factor governing the curvature structure, thereby linking relativistic kinematics directly with the underlying quantum statistics. This interplay opens new avenues for exploring geometric aspects of relativistic statistical mechanics.

In summary, our objectives are twofold. First, we construct a unified relativistic formalism for classical and quantum distributions in arbitrary spatial dimensions. Second, we employ the framework of thermodynamic geometry to extract and interpret the geometric signatures of these statistical regimes. This study provides deeper insight into the equilibrium properties of relativistic systems and contributes to the understanding of physical situations wherein relativistic and quantum effects are inherently intertwined.

The paper is organized as follows. Section \ref{2} presents the relativistic density of states and some thermodynamic quantities namely the total particle number and internal energy. Section \ref{3} develops the thermodynamic geometry and thermodynamic curvature. In section \ref{4} and \ref{5} we consider the two and three spatial dimensional systems in the relativistic regime. The relativistic condensation temperature is investigated in section \ref{6}. Finally we conclude the paper in section \ref{7}.

\section{Density of states}\label{2}

In quantum statistical mechanics, particles are fundamentally classified according to their intrinsic spin, which determines the statistical laws governing their collective behavior. Particles with integer spin, known as bosons, obey Bose-Einstein statistics and are not subject to restrictions on the number of particles occupying a single quantum state. In contrast, particles with half-integer spin, called fermions, follow Fermi-Dirac statistics, where the Pauli exclusion principle limits the occupation to at most one particle per quantum state. Despite their distinct physical implications, both types of statistics admit a unified formal description. The equilibrium occupation number of a single-particle state with energy $\epsilon$ is given by
\begin{equation}
n(\epsilon) = \frac{1}{e^{(\epsilon-\mu)/(k_B T)} + a},
\end{equation}
where $\mu$ is the chemical potential, $k_B$ is Boltzmann's constant, and $T$ denotes the temperature. The parameter $a$ specifies the quantum statistics: $a=-1$ corresponds to Bose-Einstein statistics, whereas $a=+1$ corresponds to Fermi-Dirac statistics, while $a=0$ represents the classical Maxwell-Boltzmann statistics. This compact formulation enables a unified treatment of quantum gases with different exchange symmetries.



We consider an ideal gas with particles in relativistic regime. Therefore, the energy of a particle with rest mass $m$ and momentum $p$ is
\begin{equation}\label{dispersion}
\epsilon(p) = \sqrt{p^2 c^2 + m^2 c^4},
\end{equation}
where $c$ is the speed of light. This expression naturally reduces to the familiar limits. 
In non-relativistic limit for $p \ll mc$, one finds $\epsilon \simeq mc^2 + {p^2}/{2m}$,
and in ultra-relativistic limit, for $p \gg mc$, one finds $\epsilon \simeq pc$.

The relativistic dispersion relation has important consequences for thermodynamic quantities such as the density of states, internal energy, and fluctuations.

We derive the single-particle density of states (DOS) for a relativistic dispersion relation in arbitrary dimension $D$. 
We start from the relativistic energy-momentum relation in Eq.  (\ref{dispersion}).
We note that the spectrum has a minimum value at $\epsilon=mc^{2}$, which corresponds to a particle at rest.

The density of states is obtained by counting how many single-particle momentum states fall into a given energy interval. In $D$ spatial dimensions, the number of quantum states within a spherical momentum shell between $p$ and $p+dp$ is
\begin{equation}
g(p)\,dp = \frac{V}{(2\pi\hbar)^{D}}\, S_{D}\, p^{D-1}\, dp,
\end{equation}
where $V$ is the system volume and $(2\pi\hbar)^{D}$ is the phase-space cell volume in quantum mechanics. The factor $S_{D}\,p^{D-1}$ comes from the surface area of a $D$-dimensional sphere of radius $p$, which counts how many momentum states have magnitude $p$. The surface area of the $D$-dimensional unit sphere is
\begin{equation}
S_{D} = \frac{2\pi^{D/2}}{\Gamma(D/2)}.
\end{equation}

Since the density of states is usually expressed as a function of energy, we now convert it from momentum $p$ to energy $\epsilon$. The relativistic dispersion relation gives:
\begin{equation}
p(\epsilon) = \frac{1}{c}\sqrt{\epsilon^{2}-m^{2}c^{4}},
\qquad \epsilon \ge mc^{2},
\end{equation}
which vanishes at the threshold energy $\epsilon=mc^{2}$ and increases monotonically with $\epsilon$. To perform the change of variables from momentum to energy, we need the Jacobian $dp/d\epsilon$:
\begin{equation}
\frac{dp}{d\epsilon} = \frac{\epsilon}{c\,\sqrt{\epsilon^{2}-m^{2}c^{4}}}.
\end{equation}

Combining the expressions for $p(\epsilon)$ and $dp/d\epsilon$, we obtain the crucial factor connecting the momentum-space volume element to an energy-dependent expression:
\begin{equation}
p^{D-1}\frac{dp}{d\epsilon}
=\frac{\epsilon}{c^{D}}
\left(\epsilon^{2}-m^{2}c^{4}\right)^{\frac{D-2}{2}}.
\end{equation}

By definition, the density of states $\Omega(\epsilon)$ satisfies
\begin{equation}
\Omega(\epsilon)\, d\epsilon = g(p)\, dp,
\end{equation}
The $D$-dimensional integral representation of the density of states is:

\begin{equation}
\Omega(\epsilon) = g_s \,\frac{V}{(2\pi \hbar)^D} 
\int d^D p \, \delta\!\big(\epsilon - \sqrt{p^2 c^2 + m^2 c^4}\big),
\end{equation}

Using $D$-dimensional spherical coordinates and performing the integral over the angular variables yields:
\begin{equation}
\Omega(\epsilon) = \frac{V}{(2\pi \hbar)^D}\,
\frac{2\pi^{D/2}}{\Gamma(D/2)} \,
p(\epsilon)^{D-1} \frac{dp}{d\epsilon}.
\end{equation} 
Using the expression for $g(p)$ together with the Jacobian $dp/d\epsilon$, we finally find that:

\begin{equation}\label{Dos.non.int}
\Omega(\epsilon)
=
\frac{V}{(2\pi\hbar)^D}\,\frac{2\pi^{D/2}}{\Gamma\!\big(\tfrac{D}{2}\big)}
\,
\frac{\epsilon\big(\epsilon^2-m^2c^4\big)^{\frac{D-2}{2}}}{c^{D}}.
\end{equation}

This function $\Omega(\epsilon)$ is used in the calculation of thermodynamic quantities such as total particle number and internal energy.

\begin{widetext}
\subsection{Thermodynamic Quantities}\label{Thermodynamic Quantities} 

In this subsection, using the density of state, we obtain the thermodynamic quantities of the total number of particles and the internal energy, Using the density of states and the distribution function \(n(\epsilon)\), the total number of particles and the internal energy are given by:

\begin{equation}
\label{N}N=\int_{mc^2}^{\infty} d\epsilon \; \Omega(\epsilon)\, n(\epsilon)
=
\frac{V}{(2\pi\hbar)^D}\,\frac{2\pi^{D/2}}{\Gamma\!\big(\tfrac{D}{2}\big)}\frac{1}{c^D}
\int_{mc^2}^{\infty}\! \; 
\frac{\epsilon\big(\epsilon^2-m^2c^4\big)^{\frac{D-2}{2}}}{\;z^{-1}e^{\beta\epsilon}+ a\;}d\epsilon\ .
\end{equation}
\begin{equation}
\label{U}U=\int_{mc^2}^{\infty} d\epsilon \; \Omega(\epsilon)\, \epsilon\, n(\epsilon)
=
\frac{V}{(2\pi\hbar)^D}\,\frac{2\pi^{D/2}}{\Gamma\!\big(\tfrac{D}{2}\big)}\frac{1}{c^D}
\int_{mc^2}^{\infty}\!
\frac{\epsilon^{2}\big(\epsilon^2-m^2c^4\big)^{\frac{D-2}{2}}}
{\;z^{-1}e^{\beta\epsilon}+a\;} \; d\epsilon\ .
\end{equation}
These integrals naturally account for relativistic kinematics behavior.
where, for simplicity, we set $A = 1$. Integrals in Eqs. (\ref{N}) and (\ref{U}) are calculated numerically for a specified values of $m$, $z$ and $D$. However, it is obvious that in two spatial dimension, the density of states has a simple linear relation with energy and therefore, the integrals are calculated analytically. In the following we construct the thermodynamic geometry and obtain the metric elements in two and three spatial dimension.
\end{widetext}

\section{Thermodynamic geometry}\label{3}
Ruppeiner and Weinhold introduced thermodynamic geometry as a powerful framework for analyzing the structure and behavior of thermodynamic systems from a geometric perspective~\cite{ruppeiner1979thermodynamics,weinhold1975metric}. In this approach, the space of thermodynamic parameters is treated as a Riemannian manifold, where geometric quantities such as curvature offer insights into the nature of microscopic interactions.

The Ruppeiner metric is defined as the negative Hessian of entropy with respect to extensive variables such as internal energy, volume, and particle number~\cite{ruppeiner1979thermodynamics}. In contrast, Weinhold proposed a metric based on the second derivatives of the internal energy with respect to the extensive variables~\cite{weinhold1975metric}. These two metrics are related by a conformal transformation involving the system temperature~\cite{salamon1984relation}.

Legendre transformations facilitate the transition between different thermodynamic potentials—such as the Helmholtz and Gibbs free energies—by exchanging the roles of extensive and intensive variables. In addition to these well known metric formalism, the Fisher–Rao information metric emerges naturally within the framework of information geometry. It is constructed from the second derivatives of the logarithm of the partition function with respect to non-extensive parameters, typically the inverse temperature \(\beta = 1 / k_B T\) and the dimensionless chemical potential parameter \(\gamma = -\mu / k_B T\)~\cite{ruppeiner1995riemannian,janyszek1990riemannian,brody1995geometrical,amari2000methods}:
\begin{equation}
g_{ij} = \partial_i \partial_j \ln \mathcal{Z},
\end{equation}
where \(\partial_i\) denotes partial differentiation with respect to the \(i\)-th non-extensive thermodynamic variable, and \(\mathcal{Z}\) is the partition function. For classical and quantum ideal gases, the logarithm of the partition function typically depends on \(\beta\) and \(\gamma\), with the system volume held fixed. Consequently, the geometry of thermodynamic fluctuations can often be effectively described in a two-dimensional parameter space.

The Christoffel symbols, which act as connection coefficients, are expressed in terms of the metric tensor components as \cite{schutz1980}:
\begin{equation}
\Gamma^i_{jk} = \frac{1}{2} g^{im} \left( \partial_k g_{mj} + \partial_j g_{mk} - \partial_m g_{jk} \right),
\end{equation}
where \(g^{im}\) denotes the components of the inverse metric tensor, and \(\partial_k g_{ij}\) represents the partial derivative of the metric component with respect to the parameter \(k\).

The Riemann curvature tensor, which encodes the intrinsic curvature of the thermodynamic manifold, is defined by:
\begin{equation}
R^i_{jkl} = \partial_k \Gamma^i_{lj} - \partial_l \Gamma^i_{kj} + \Gamma^i_{km} \Gamma^m_{lj} - \Gamma^i_{lm} \Gamma^m_{kj}.
\end{equation}
From the Riemann tensor, the Ricci tensor, a second-rank contraction, is obtained as:
\begin{equation}
R_{ij} = R^m_{imj},
\end{equation}
and the scalar curvature, often referred to as the thermodynamic scalar curvature, is given by:
\begin{equation}
R = g^{ij} R_{ij}.
\end{equation}

In the case of a two-dimensional thermodynamic parameter space, the Ricci scalar simplifies significantly.

The system under consideration involves two fluctuating thermodynamic parameters; therefore, the dimension of the thermodynamic parameter space is equal to two. 
Janyszek and Mrugala demonstrated that when the metric components are expressed exclusively as second derivatives of a specific thermodynamic potential, the thermodynamic curvature can be expressed in terms of second and third derivatives. Furthermore, the sign convention used here is consistent with that adopted by Janyszek and Mrugala. For a two-dimensional thermodynamic parameters space spanned by the parameters \( \beta \) and \( \gamma \), the Ricci scalar can be written as:

\begin{equation}
R = - \frac{
\begin{vmatrix}
g_{\beta\beta} & g_{\beta\gamma} & g_{\gamma\gamma} \\
 g_{\beta\beta,\beta} & g_{\beta\gamma,\beta} & g_{\gamma\gamma,\beta} \\
 g_{\beta\beta,\gamma} & g_{\beta\gamma,\gamma} &  g_{\gamma\gamma,\gamma}
\end{vmatrix}
}{2
\begin{vmatrix}
g_{\beta\beta} & g_{\beta\gamma} \\
g_{\beta\gamma} & g_{\gamma\gamma}
\end{vmatrix}^2}.
\end{equation}

Thermodynamic curvature serves as a key tool to probe statistical interactions. An ideal gas with particles that obeying classical Maxwell-Boltzmann statistics and non-relativistic dispersion relation corresponds to a flat thermodynamic space with zero curvature, indicating neutral interactions. For ideal quantum gases with non-relativistic particles, the curvature is positive for bosons (attractive statistical interactions) and negative for fermions (repulsive statistical interactions) \cite{janyszek1990,ruppeiner1995riemannian}. Moreover, singularities in thermodynamic curvature indicate phase transitions. For instance, the condensation point of an ideal boson gas, at fugacity $z=1$, corresponds to a divergence in the curvature. This approach has been applied to generalized statistics, including deformed, non-extensive, and dual statistics, showing that the singular points of the curvature coincide with the systems' phase transition points.The thermodynamic geometry of various generalized statistics has also been investigated, including fractional exclusion, Polychronakos, Gentile, deformed, Kaniadakis, nonextensive,alpha and some other generalized statistics \cite{mirza2010thermodynamic,mirzathermodynamic2011,oshima1999riemann,mirza2011condensation,adli2019nonperturbative,mohammadzadeh2017thermodynamic,mehri2020thermodynamic,mohammadzadeh2016perturbative,ebadi2022thermodynamic,alphastatistics,seifi2025mittag,seifi2025intrinsic}.

To evaluate the thermodynamic curvature in relativistic regime, the metric elements of the thermodynamic parameters space are first calculated using the expressions for the total particle number and internal energy. Partial derivatives are taken with respect to the thermodynamic parameters \(\beta\) and \(\gamma\).  \cite{ruppeiner1979thermodynamics, ruppeiner1995riemannian}. The metric elements, referred to as the Fisher-Rao metric, are calculated as follows:
\begin{widetext}
\begin{align}
g_{\beta\beta} &= \frac{\partial^2 \ln \mathcal{Z}}{\partial \beta^2} 
= -\left(\frac{\partial U}{\partial \beta}\right)_{\gamma}= 
\frac{V}{(2\pi\hbar)^D}
\frac{2\pi^{D/2}}{\Gamma(D/2)}
\frac{1}{c^D}
\int_{mc^2}^{\infty}
\frac{
\epsilon^{3}
\left(\epsilon^2-m^2c^4\right)^{\frac{D}{2}-1}
\, z^{-1} e^{\beta\epsilon}
}
{\left(z^{-1}e^{\beta\epsilon}+a\right)^2}
\, d\epsilon ,
\label{gbb}
\\[2mm]
g_{\beta\gamma} &= g_{\gamma\beta} 
= \frac{\partial^2 \ln \mathcal{Z}}{\partial \beta \partial \gamma} 
= -\left(\frac{\partial N}{\partial \beta}\right)_{\gamma}=
\frac{V}{(2\pi\hbar)^D}
\frac{2\pi^{D/2}}{\Gamma(D/2)}
\frac{1}{c^D}
\int_{mc^2}^{\infty}
\frac{
\epsilon^{2}
\left(\epsilon^2-m^2c^4\right)^{\frac{D}{2}-1}
\, z^{-1} e^{\beta\epsilon}
}
{\left(z^{-1}e^{\beta\epsilon}+a\right)^2}
\, d\epsilon ,
\label{gbg}
\\[2mm]
g_{\gamma\gamma} &= \frac{\partial^2 \ln \mathcal{Z}}{\partial \gamma^2} 
= -\left(\frac{\partial N}{\partial \gamma}\right)_{\beta}=
\frac{V}{(2\pi\hbar)^D}
\frac{2\pi^{D/2}}{\Gamma(D/2)}
\frac{1}{c^D}
\int_{mc^2}^{\infty}
\frac{
\epsilon
\left(\epsilon^2-m^2c^4\right)^{\frac{D}{2}-1}
\, z^{-1} e^{\beta\epsilon}
}
{\left(z^{-1}e^{\beta\epsilon}+a\right)^2}
\, d\epsilon .
\label{ggg}
\end{align}

By differentiating Eqs.~(\ref{gbb}), (\ref{gbg}), and (\ref{ggg}) with respect to \( \beta \) and \( \gamma \), we obtain the required terms to evaluate the thermodynamic curvature as follows:
\begin{gather}
g_{\beta\beta,\beta} = \frac{\partial g_{\beta\beta}}{\partial \beta} 
= -\int_{mc^2}^{\infty} 
\frac{2\,\epsilon^4 (\epsilon^2-m^2c^4)^{D/2-1} e^{2\beta\epsilon}}{(e^{\beta\epsilon} z^{-1}+a)^3 z^2} 
- \frac{\epsilon^4 (\epsilon^2-m^2c^4)^{D/2-1} e^{\beta\epsilon}}{(e^{\beta\epsilon} z^{-1}+a)^2 z} \, d\epsilon
\\[2mm]
g_{\beta\gamma,\beta} = \frac{\partial g_{\beta\gamma}}{\partial \beta} 
= -\int_{mc^2}^{\infty} 
\frac{2\,\epsilon^3 (\epsilon^2-m^2c^4)^{D/2-1} e^{2\beta\epsilon}}{(e^{\beta\epsilon} z^{-1}+a)^3 z^2} 
- \frac{\epsilon^3 (\epsilon^2-m^2c^4)^{D/2-1} e^{\beta\epsilon}}{(e^{\beta\epsilon} z^{-1}+a)^2 z} \, d\epsilon
\\[2mm]
g_{\beta\gamma,\gamma} = \frac{\partial g_{\beta\gamma}}{\partial \gamma} 
= -\int_{mc^2}^{\infty} 
\frac{2\,\epsilon^2 (\epsilon^2-m^2c^4)^{D/2-1} e^{2\beta\epsilon}}{(e^{\beta\epsilon} z^{-1}+a)^3 z^2} 
+ \frac{\epsilon^2 (\epsilon^2-m^2c^4)^{D/2-1} e^{\beta\epsilon}}{(e^{\beta\epsilon} z^{-1}+a)^2 z} \, d\epsilon
\\[2mm]
g_{\gamma\gamma,\gamma} = \frac{\partial g_{\gamma\gamma}}{\partial \gamma} 
= - \int_{mc^2}^{\infty} 
\frac{\epsilon (\epsilon^2-m^2c^4)^{D/2-1} e^{\beta\epsilon}}{(e^{\beta\epsilon} z^{-1}+a)^2 z} \, d\epsilon
- \int_{mc^2}^{\infty} 
\frac{2\,\epsilon (\epsilon^2-m^2c^4)^{D/2-1} e^{2\beta\epsilon}}{(e^{\beta\epsilon} z^{-1}+a)^3 z^2} 
- \frac{2\,\epsilon (\epsilon^2-m^2c^4)^{D/2-1} e^{\beta\epsilon}}{(e^{\beta\epsilon} z^{-1}+a)^2 z} \, d\epsilon
\end{gather}
\end{widetext}

For convenience, all prefactors outside the integrals have been absorbed into the constant \(A\), and in what follows we take \(A = 1\). Using the derived relations, all the necessary expressions are now available to determine the thermodynamic curvature and analyze the system's behavior of relativistic particles.
\section{Relativistic Statistical Behavior in Two Spatial Dimension}\label{4}

In this section, we analyze the statistical behavior of relativistic particles obeying Maxwell-Boltzmann, Bose-Einstein, and Fermi-Dirac statistics in two spatial dimensions. An important advantage of the two-dimensional case is that the relevant thermodynamic quantities can be obtained in closed analytical form. This allows for an exact evaluation of the thermodynamic metric and curvature, providing a transparent benchmark for comparison with higher-dimensional cases.

Using the relativistic density of states in $D=2$, the total particle number and internal energy can be expressed in terms of polylogarithm functions as
\begin{widetext}
\begin{equation}
N=\frac{V}{2\pi \hbar^2 c^2 \beta^2}\frac{1}{a}\left[
\beta m c^2 \ln\!\left(1+a z e^{-\beta m c^2}\right)
+\operatorname{Li}_2\!\left(-a z e^{-\beta m c^2}\right)
\right],
\end{equation}
\begin{equation}
U=\frac{V}{2\pi \hbar^2 c^2 \beta^3}\frac{1}{a}\left[
(\beta m c^2)^2 \ln\!\left(1+a z e^{-\beta m c^2}\right)
+2(\beta m c^2)\operatorname{Li}_2\!\left(-a z e^{-\beta m c^2}\right)
+2\operatorname{Li}_3\!\left(-a z e^{-\beta m c^2}\right)
\right].
\end{equation}
\end{widetext}

The internal energy and total particle number are explicit functions of $\beta$ and $z$. The derivatives with respect to $\beta$, required for the evaluation of the thermodynamic metric elements, can be computed straightforwardly. For derivatives with respect to the fugacity $z$, we use the standard polylogarithm identity
$x\,\partial_x \mathrm{Li}_n(x)=\mathrm{Li}_{n-1}(x)$.
Due to the length of the resulting expressions, the explicit forms of the metric elements and their derivatives are not presented here.

\begin{figure}[h]
    \centering
    \includegraphics[width=0.45\textwidth]{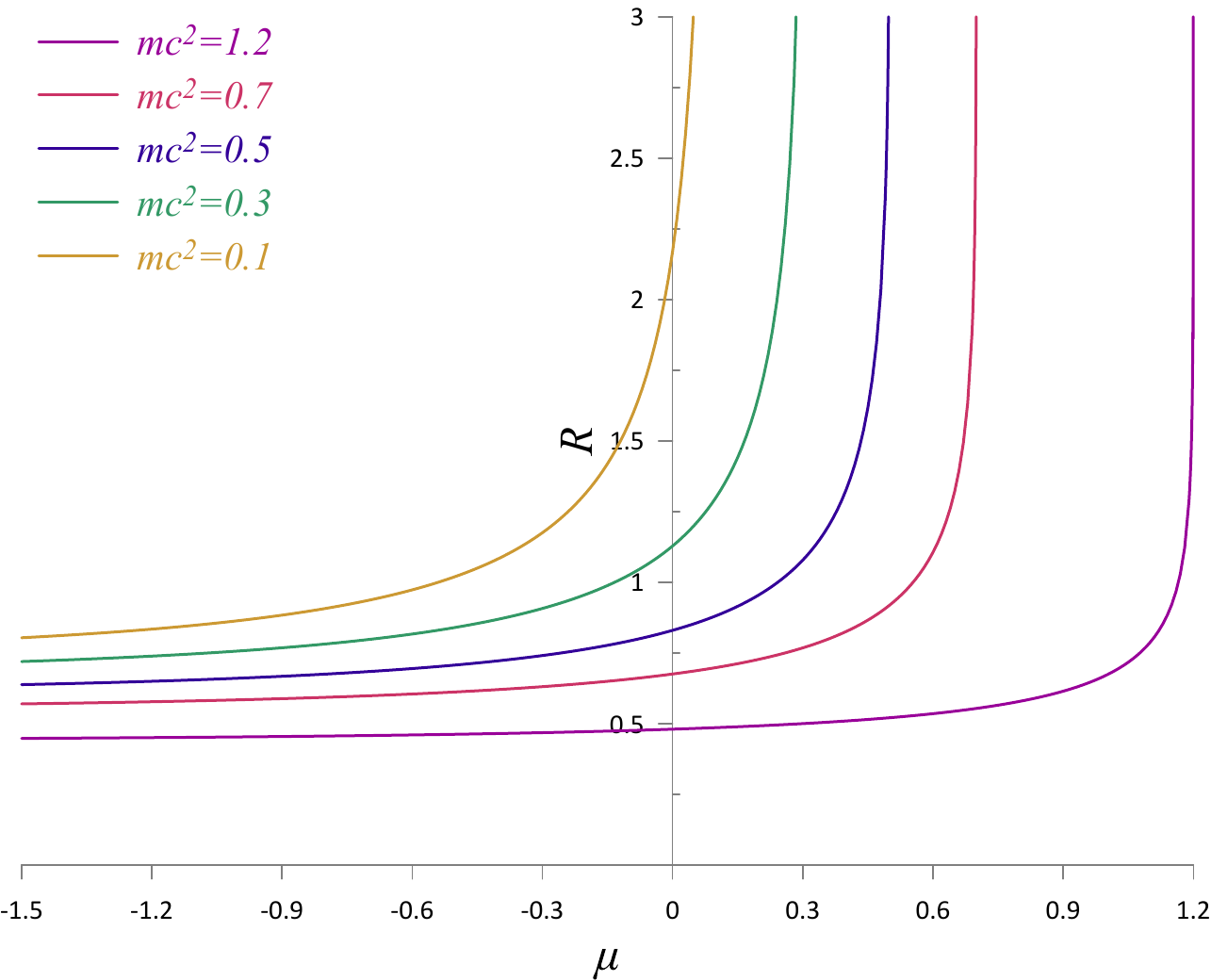}
    \caption{Thermodynamic curvature of bosons as a function of the chemical potential for a two-dimensional system and isothermal process $(\beta = 1)$  in relativistic regime.}
    \label{fig:B2}
\end{figure}

\begin{figure}[h]
    \centering
    \includegraphics[width=0.45\textwidth]{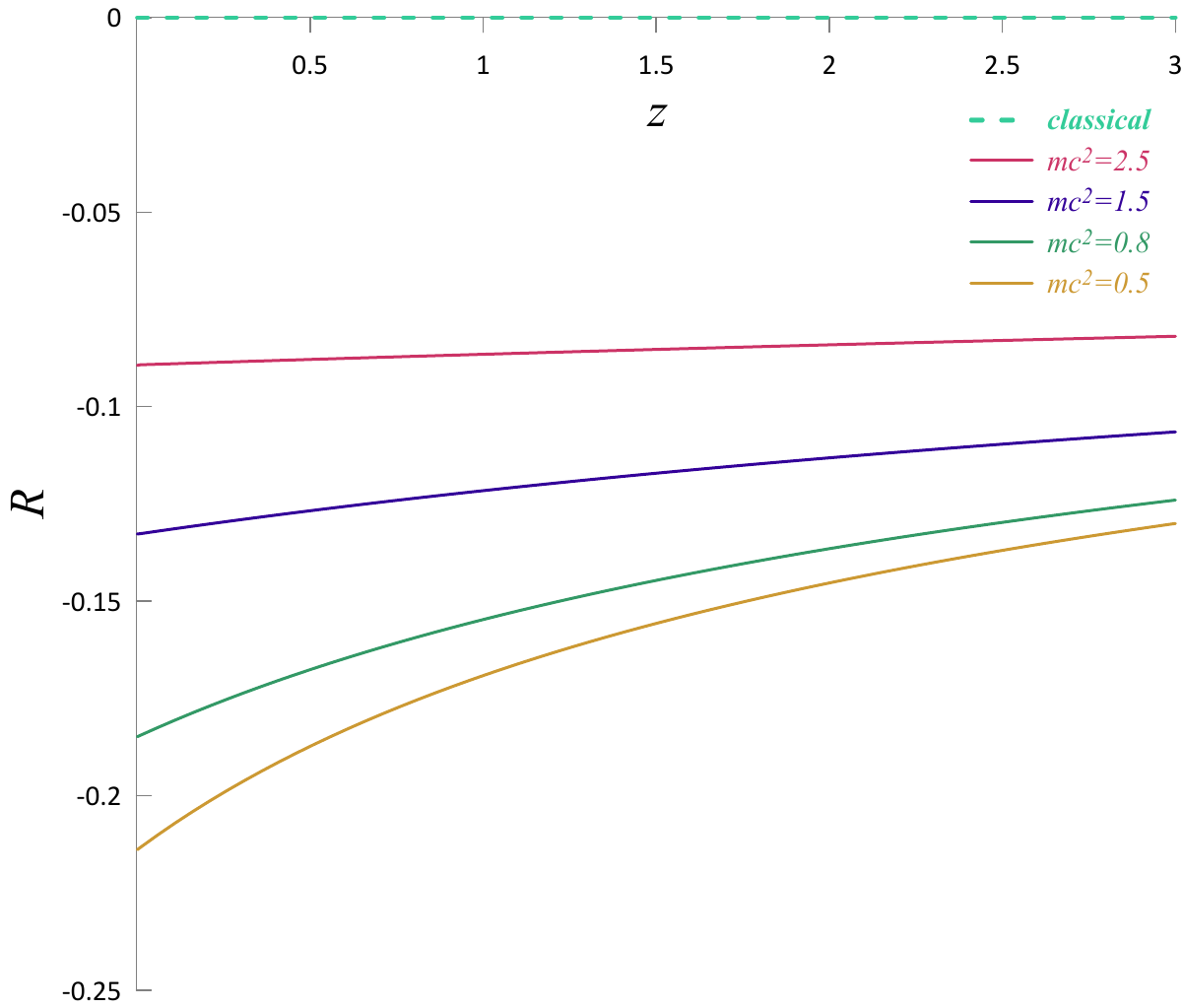}
    \caption{Thermodynamic curvature of fermions as a function of the chemical potential for a two-dimensional system and isothermal process $(\beta = 1)$  in relativistic regime.}
    \label{fig:F2}
\end{figure}

The thermodynamic curvature $R$, evaluated exactly from these analytical expressions, is shown as a function of the fugacity in Figs.~\ref{fig:B2} and \ref{fig:F2} for bosons and fermions, respectively. The results indicate that the qualitative behavior of the thermodynamic curvature in the two-dimensional relativistic regime closely resembles that of the three-dimensional case. In particular, $R$ remains positive for Bose-Einstein statistics and negative for Fermi-Dirac statistics over the entire physical range of fugacity.

As in the non-relativistic case, the thermodynamic curvature is positive (negative) for Bose-Einstein (Fermi-Dirac) statistics, reflecting effective attractive (repulsive) statistical interactions. The main difference arises in the location of the critical point. While in the non-relativistic regime the curvature diverges at $\mu=0$ (or equivalently $z=1$), in the relativistic regime the singularity occurs at the critical chemical potential $\mu_c = m c^2$.

\section{Relativistic Statistical Behavior in three Spatial Dimension}\label{5}

We now turn to the three-dimensional relativistic regime, where the thermodynamic quantities and the associated geometric structures cannot, in general, be obtained in closed analytical form. In this case, the thermodynamic curvature is evaluated numerically from the integral expressions derived in the previous sections.

Within the framework of thermodynamic geometry, we investigate the behavior of particles obeying Maxwell-Boltzmann, Bose-Einstein, and Fermi-Dirac statistics by analyzing the thermodynamic curvature $R$ as a function of the fugacity $z$. This geometric quantity provides direct insight into the nature of effective statistical interactions, while divergences of $R$ signal the presence of phase transitions.

\begin{figure}[h]
    \centering
    \includegraphics[width=0.45\textwidth]{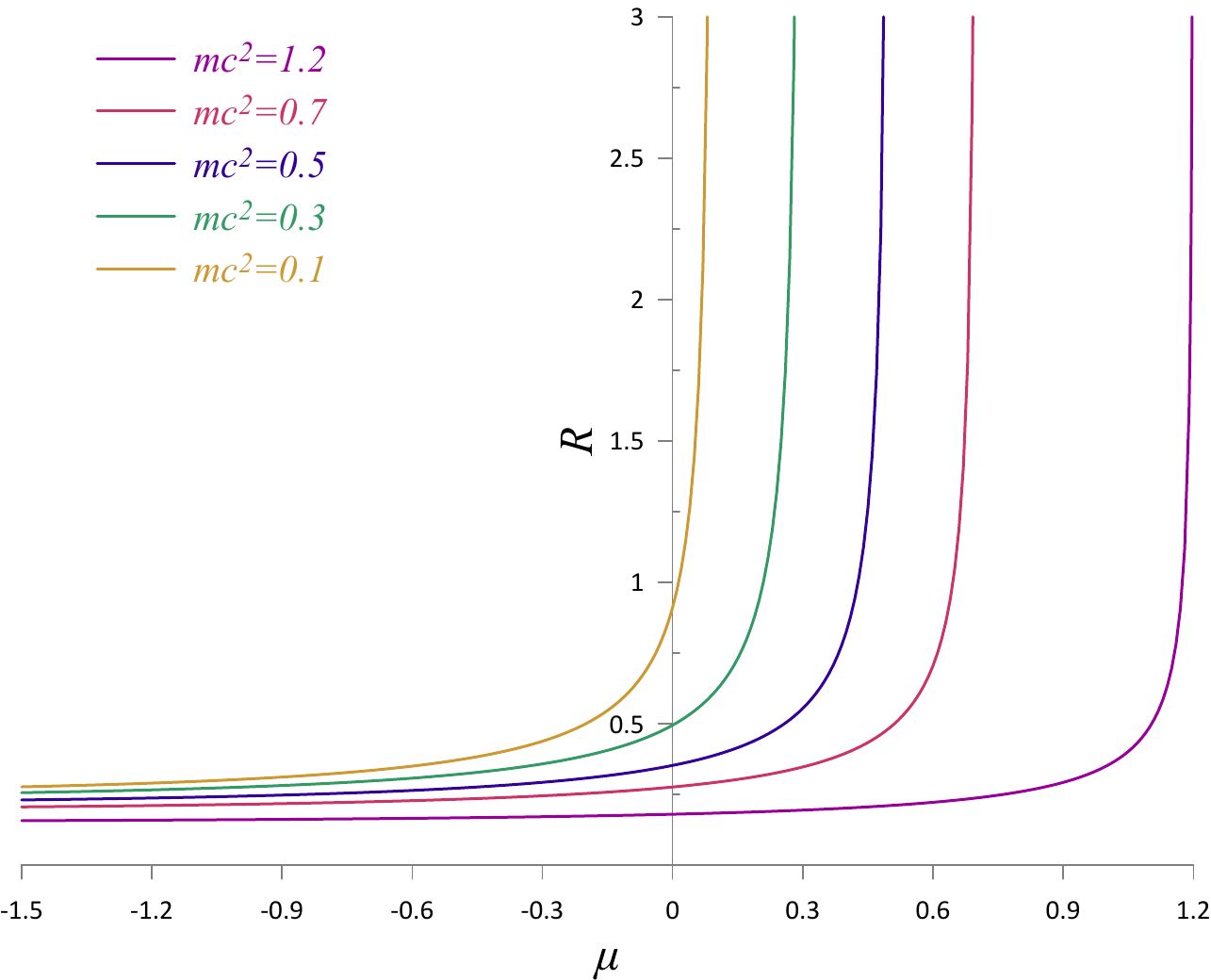}
    \caption{Thermodynamic curvature of bosons as a function of the chemical potential for a three-dimensional system and isothermal process $(\beta = 1)$  in relativistic regime.}
    \label{fig:B3}
\end{figure}

\begin{figure}[h]
    \centering
    \includegraphics[width=0.45\textwidth]{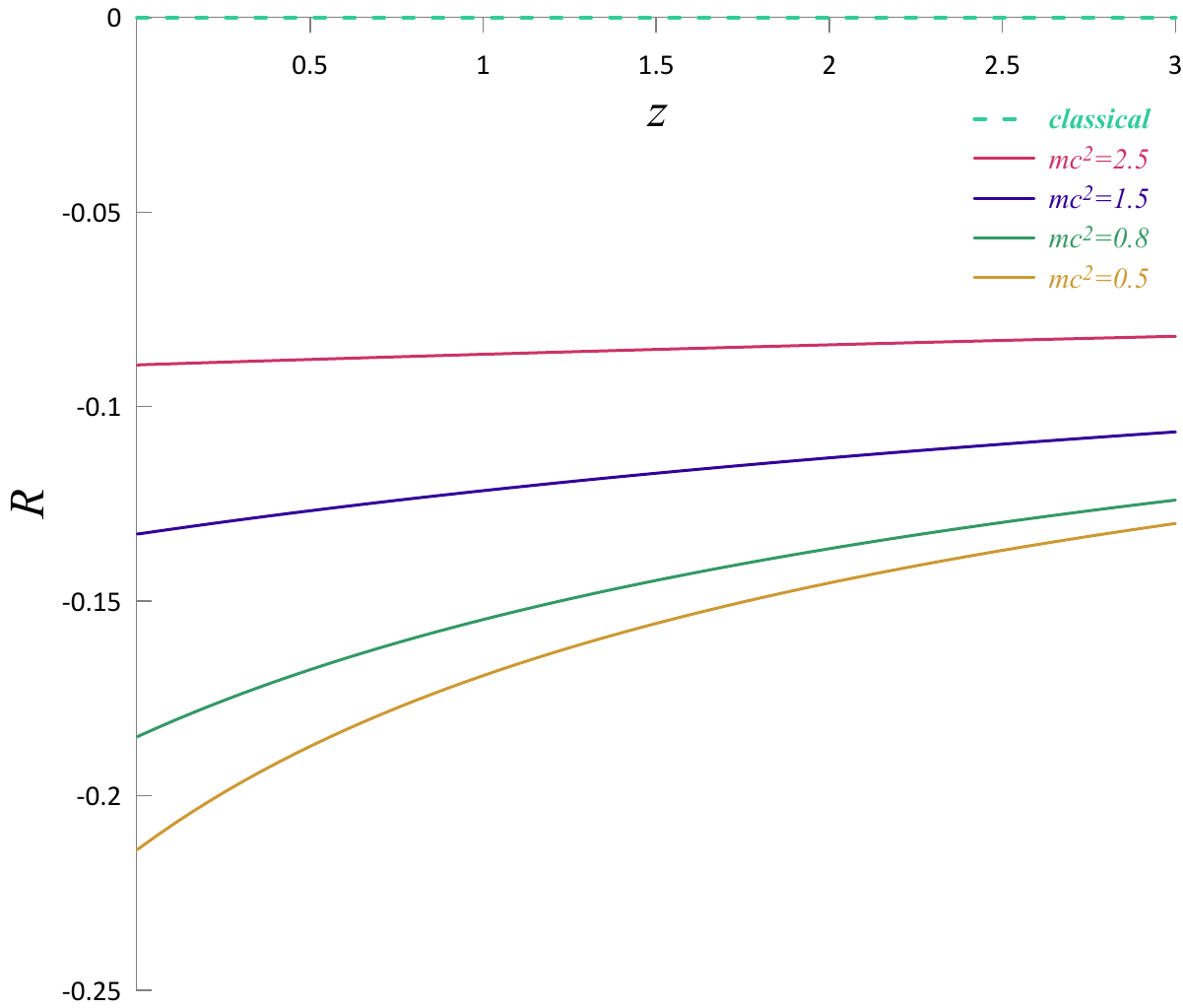}
    \caption{Thermodynamic curvature of fermions as a function of the chemical potential for a three-dimensional system and isothermal process $(\beta = 1)$  in relativistic regime.}
    \label{fig:F3}
\end{figure}

As shown in Figs.~\ref{fig:B3} and \ref{fig:F3}, for Bose-Einstein statistics $(a=-1)$ the thermodynamic curvature is always positive $(R>0)$, indicating effective attractive interactions among bosons. In contrast, for Fermi-Dirac statistics $(a=1)$ the curvature remains negative $(R<0)$ over the entire physical range, reflecting effective repulsive interactions originating from the Pauli exclusion principle. For Maxwell-Boltzmann statistics $(a=0)$, the thermodynamic curvature vanishes identically $(R=0)$, corresponding to a flat thermodynamic geometry characteristic of a classical ideal gas.

In the relativistic regime, increasing the particle mass drives both bosonic and fermionic systems toward the classical Maxwell-Boltzmann limit. A notable relativistic effect is the shift of the critical fugacity associated with Bose-Einstein condensation. Unlike the non-relativistic case, where condensation occurs at $z_c=1$, the relativistic condensation point is given by
$z_c = e^{\beta m c^2}$.
At this value, the denominator of the Bose-Einstein distribution diverges for the ground state, signaling the onset of condensation.

To maintain generality and avoid fixing the system at a particular temperature, we do not present the bosonic thermodynamic curvature solely as a function of the fugacity. Instead, the curvature is plotted as a function of the chemical potential $\mu$, which is physically restricted to the range $-\infty < \mu \leq m c^2$. At the critical chemical potential $\mu_c = m c^2$, the thermodynamic curvature diverges, clearly indicating the Bose-Einstein condensation transition, as shown in Fig.~\ref{fig:B3}. For fermions, the curvature is plotted directly as a function of the fugacity, as shown in Fig.~\ref{fig:F3}.

\section{Condensation Temperature in the Relativistic Regime}\label{6}

In this section, we derive the condensation temperature of an ideal relativistic Bose gas, emphasizing its dependence on the rest mass $m$. Of particular interest is the small-mass limit, where significant deviations from the classical (non-relativistic) behavior emerge.

At the critical point of Bose–Einstein condensation, the chemical potential reaches its maximum allowable value. We showed in previous section that the thermodynamic curvature is singular at $\mu_c = mc^2$. At $T = T_c$ and $\mu = mc^2$, the total number density $n = N/V$ takes the exact form

\begin{equation}
n = \frac{1}{2\pi^2\hbar^3} \int_0^\infty dp\, p^2 \frac{1}{\exp\Big[\dfrac{\sqrt{p^2c^2 + m^2c^4} - mc^2}{k_B T_c}\Big] - 1}.
\end{equation}

To evaluate this integral exactly, it is convenient to introduce a dimensionless variable that measures the excitation energy in units of the rest energy:

\begin{equation}\label{n1}
y = \frac{E - mc^2}{mc^2} = \frac{\sqrt{p^2c^2 + m^2c^4} - mc^2}{mc^2},
\end{equation}

or equivalently $E = mc^2(1+y)$. The momentum then becomes:

\begin{equation}
p = mc\sqrt{y(y+2)}, \qquad dp = mc\frac{y+1}{\sqrt{y(y+2)}}dy.
\end{equation}

Substituting these expressions into Eq.~(\ref{n1}) yields an exact reformulation:

\begin{equation}\label{n2}
\frac{n\hbar^3}{m^3c^3} = \frac{1}{2\pi^2} \int_0^\infty dy\, \frac{(y+1)\sqrt{y(y+2)}}{e^{\frac{mc^2}{k_B T_c} y} - 1}.
\end{equation}




In the non-relativistic limit, the condensation temperature is obtained from the following relation

\begin{equation}
\frac{n\hbar^3}{m^3c^3} = \zeta\!\left(\frac{3}{2}\right) 
\left( \frac{k_B T_c}{mc^2} \right)^{3/2}, 
\end{equation}
where, $\zeta(x)$ denoted the Riemann zeta function. 

\begin{figure}[htbp]
    \centering
    \includegraphics[width=0.45\textwidth]{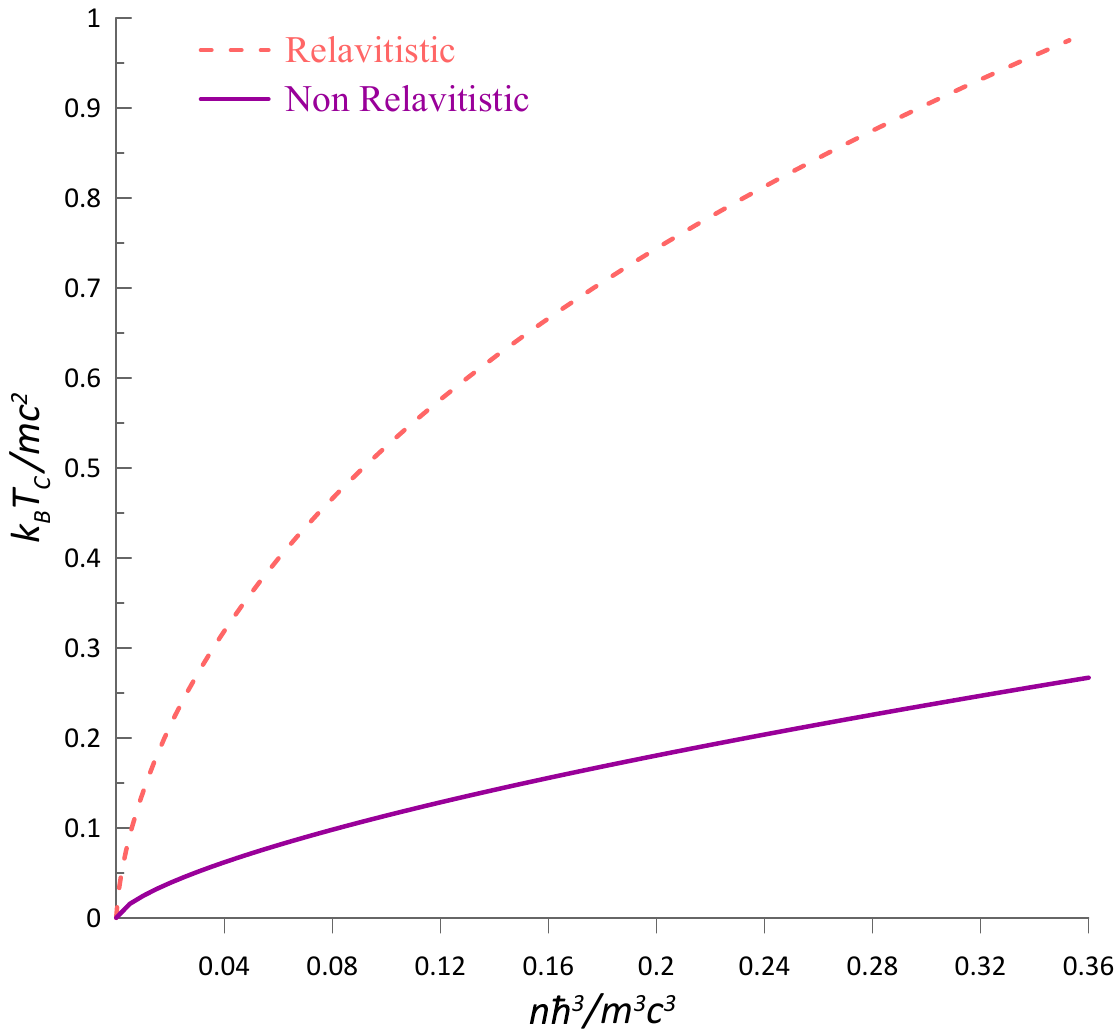}
    \caption{Critical condensation temperature; $T_c$ as a function of dimensionless number density for relativistic (dashed) and non-relativistic (solid) Bose gases.}
    \label{fig:TM}
\end{figure}

Figure~\ref{fig:TM} shows the ratio ${k_B T_c}/{mc^2}$ as a function of the dimensionless density ${n\hbar^3}/{m^3c^3}$ in both the relativistic and non-relativistic regimes. It is evident that for particles with masses satisfying $m \gg n^{1/3}\hbar/c$, the non-relativistic limit is recovered and the condensation temperature obeys $T_c \ll mc^2/k_B$. In this limit, the results obtained from relativistic and non-relativistic evaluations coincide. Conversely, for particles with masses much smaller than this threshold, $m \ll n^{1/3}\hbar/c$, relativistic effects become significant and relativistic corrections to the condensation temperature are essential. In this regime, the relativistic condensation temperature is substantially larger than the value predicted by the non-relativistic approximation. For example, when considering dark matter models constructed from the condensation of ultra-light bosonic fields, relativistic corrections to the condensation temperature become indispensable \cite{grether2007bose,boehmer2007can}. These corrections significantly modify the thermodynamic and dynamical characteristics of the condensate and, therefore, must be incorporated to obtain physically consistent cosmological predictions.

\section{Conclusion}\label{7}

We have investigated the thermodynamic geometry of classical and quantum ideal gases in the relativistic regime by combining relativistic statistical mechanics with the framework of information geometry. By incorporating the full relativistic energy-momentum dispersion relation into the density of states, we developed a unified formalism applicable to Maxwell-Boltzmann, Bose-Einstein, and Fermi-Dirac statistics in arbitrary spatial dimensions. The Fisher-Rao information metric, constructed from the partition function in terms of the inverse temperature and chemical potential, provided a natural geometric description of thermodynamic fluctuations in relativistic systems.

Our analysis demonstrates that the characteristic geometric signatures of quantum statistics remain robust in the relativistic domain. In both two- and three-dimensional systems, the thermodynamic curvature is found to be positive for bosonic gases and negative for fermionic gases, reflecting effective attractive and repulsive statistical interactions, respectively. In contrast, classical relativistic gases exhibit a vanishing curvature, consistent with the absence of statistical correlations. These results confirm that thermodynamic curvature continues to serve as a meaningful geometric indicator of microscopic statistical behavior even when relativistic kinematics plays a central role.

A key relativistic modification revealed by our study concerns the location of curvature singularities. While in non-relativistic quantum gases such divergences occur at vanishing chemical potential, we find that in the relativistic regime the singular behavior is shifted to a mass-dependent critical value, $\mu_{c} = mc^{2}$. This shift provides a clear geometric manifestation of relativistic kinematics and highlights the fundamental role of the particle rest mass in shaping the thermodynamic structure of relativistic systems.

The two-dimensional case admits exact analytical expressions in terms of polylogarithmic functions, allowing for a transparent and fully analytical evaluation of the thermodynamic metric and curvature. This provides a valuable benchmark for understanding relativistic effects in higher dimensions. In three spatial dimensions, where closed-form expressions are generally unavailable, numerical evaluation confirms the persistence of the same qualitative geometric features, demonstrating the consistency and robustness of the thermodynamic geometric approach across different dimensionalities.

In addition, we examined the relativistic Bose-Einstein condensation temperature and showed that relativistic kinematics introduces explicit mass-dependent corrections to the familiar non-relativistic result. In the limit of sufficiently large particle mass, the standard non-relativistic condensation temperature is smoothly recovered. However, for light and ultra-light bosonic particles, relativistic effects become significant, leading to noticeable deviations and an enhancement of the condensation temperature compared to its non-relativistic counterpart. This behavior underscores the importance of relativistic corrections in systems involving very small masses, where the energy scale associated with $mc^{2}$ becomes comparable to thermal energies.

Overall, our results highlight thermodynamic geometry as a powerful and unifying framework for probing the equilibrium properties of relativistic statistical systems. Possible extensions of the present work include the incorporation of particle interactions, generalized or deformed statistics, and curved space-time backgrounds, where local definitions of temperature and geometry may further enrich the thermodynamic structure. Such extensions could be particularly relevant for relativistic plasmas, early-universe scenarios, and systems involving ultra-light particles, where quantum and relativistic effects are intrinsically intertwined.

\bibliography{sample}

\end{document}